# Pattern Recognition of Scrap Plastic Misclassification in Global Trade Data

MUHAMMAD SUKRI BIN RAMLI
Asia School of Business
Kuala Lumpur, Malaysia
Email: m.binramli@sloan.mit.edu

*Abstract*

*Cheap, contaminated plastic scraps are often mislabelled under high-value HS codes, skewing global trade data and weakening agreements like the Basel Convention. We detect this fraud by identifying an "inverse price-volume signature" a pattern where reported volumes climb as unit prices fall. Our transparent machine-learning pipeline analyses UN Comtrade HS 39 data (2020-2024), engineers both basic and advanced price-volume measures, groups products with K-Means clustering, and applies a Random Forest model that flags high-risk segments with 93.75 % accuracy (0.89 precision, 0.92 recall). Explainable AI shows that price drops, volatile pricing, and rising volumes drive these alerts. Testing on firm-level records (2019-2025) confirms that global red flags especially for HS 390210 translate into actionable watchlists. This scalable framework equips regulators with a risk-based inspection tool under policies like Malaysia's 2025 HS 39.15 Certificate of Approval and offers data-driven support for the international Plastics Treaty.*

**Keywords**: *Misclassification, Plastic Scrap, Machine Learning, Pattern Recognition, Non-Intrusive Inspection.*

## 1. INTRODUCTION

The global trade in plastic waste has been profoundly disrupted since China's 2018 National Sword policy sharply curtailed historic import channels, displacing massive flows of scrap plastic and creating intense pressure on exporters to find new markets (INTERPOL, 2020). This pressure has fuelled a surge in illicit practices, with one of the most pervasive tactics being the deliberate misclassification of low-value, contaminated scrap under customs codes reserved for higher-value materials. This practice is more than a technical violation; it is a significant environmental crime that perpetuates "waste colonialism" by externalizing environmental harm from developed to developing nations (Liboiron, 2021). The Basel Convention provides a foundational governance structure for global plastic waste trade, and although certain operational flexibilities and enforcement dynamics have permitted continued cross-border movements, these circumstances now inform a pathway for iterative policy calibration and procedural enhancement (Basel Action Network, 2022).

**Figure 1. Illustrative Flow of Plastic Waste (HS 3915) from Top Exporters to Malaysia.**

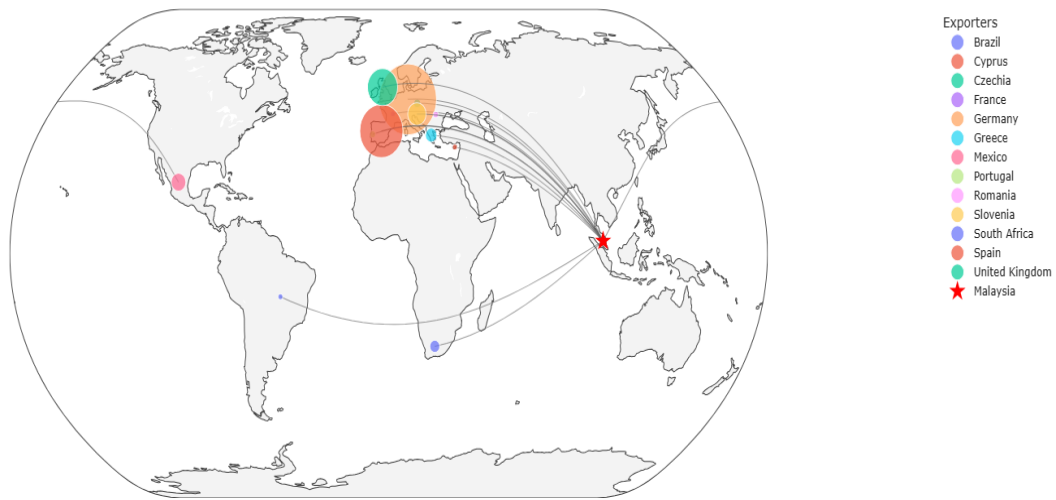

Source: Processed by Author (2025)

This paper argues that this widespread misclassification is not invisible. Instead, it leaves a detectable "inverse price-volume signature" in international trade data: a dynamic where reported trade volumes remain stable or even rise while average unit prices collapse. This pattern is inconsistent with normal commodity market behaviour and, much like a bearish signal in financial markets where falling prices and high volume indicate stress, it functions as a red flag for illicit activity. We propose that this signature is the fingerprint of a dual-motive crime. Primarily, it serves as a method of sanction evasion, allowing prohibited waste to cross borders disguised as legitimate goods. Concurrently, it mirrors classic value manipulation patterns used in Trade-Based Money Laundering (TBML), a significant financial crime (Financial Action Task Force, 2021).

**Figure 2. Comparison of General Trend of HS 39 Vs Specific HS 39 Trend**

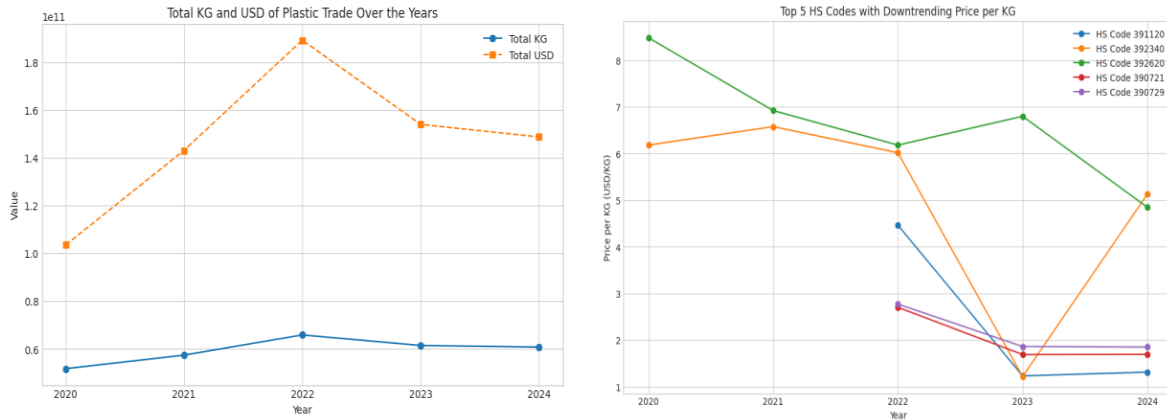

Source: Processed by Author (2025)

**Figure 3. Analogy of Inverse Price -Volume in Export Commodity Mimicking Bearish Signal in Financial Markets**

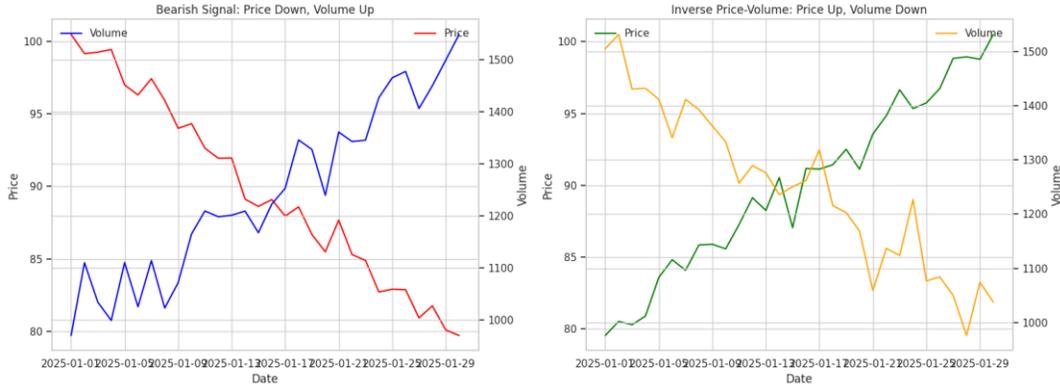

Source: Processed by Author (2025)

The dynamic tension between illicit actors and enforcement authorities can be conceptualized as a system of competing feedback loops. A reinforcing loop (R1) shows bad actors driving misclassification, which paradoxically makes the illicit fingerprint more distinct as the problem grows. This, in turn, triggers a balancing loop (B1) where the clearer signature enables better detection and enforcement, theoretically curbing the illicit activity. The challenge lies in making the balancing loop of enforcement more effective than the reinforcing loop of crime. To ground this dynamic in a real-world context, this study focuses on Malaysia.

**Figure 4. Causal Loop Diagram of Systems Dynamic Between Bad Actor and Customs Authorities on Global Trade Data.**

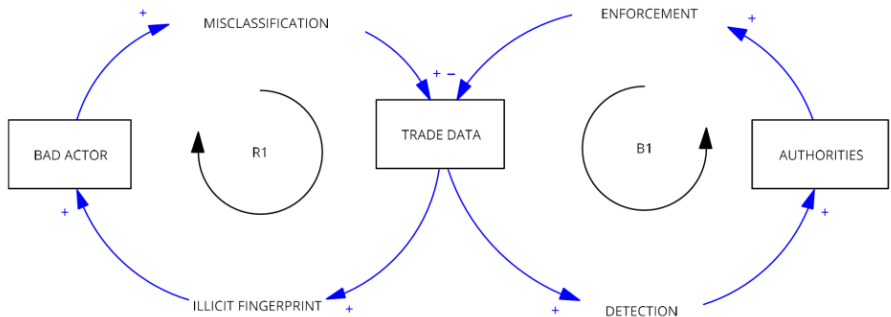

Source: Processed by Author (2025)

To investigate this dynamic, we focus on Malaysia, one of Southeast Asia's top plastic importers. The country's trade data provides a compelling case study, showing highly volatile import trends for plastic waste (HS 3915) alongside a strong market recovery driven by key European partners, particularly the United Kingdom, Germany, and Spain, which all showed significant upward trends from 2021 onwards, cementing their roles as crucial suppliers to the Malaysian market.

**Figure 5. Import Volume Trends to Malaysia Over Time for Top 15 Exporters.**

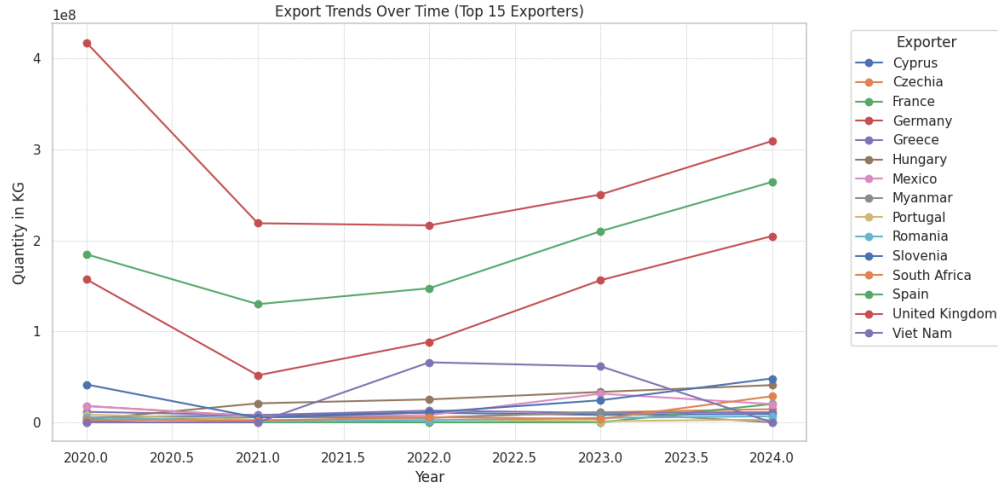

Source: Processed by Author (2025)

To move beyond simply identifying the problem, this study develops and validates a transparent, explainable machine learning framework to identify these illicit trade signatures. This report weaves together macro-level UN Comtrade data with proprietary firm-level records and a sequence of machine-learning workflows including forecasting, anomaly detection, and explainable classification. By applying K-Means clustering on features such as average volume, price volatility, and trend slopes, our analysis identifies four distinct market segments. This allows for the differentiation of high-volume and emerging commodities marked as "at-risk" from more stable, lower-risk groups, enabling regulators to prioritize inspections and allocate resources more effectively.

**Figure 6. Annual Malaysian Imports of Plastic Waste (HS 3915) by Subcategory.**

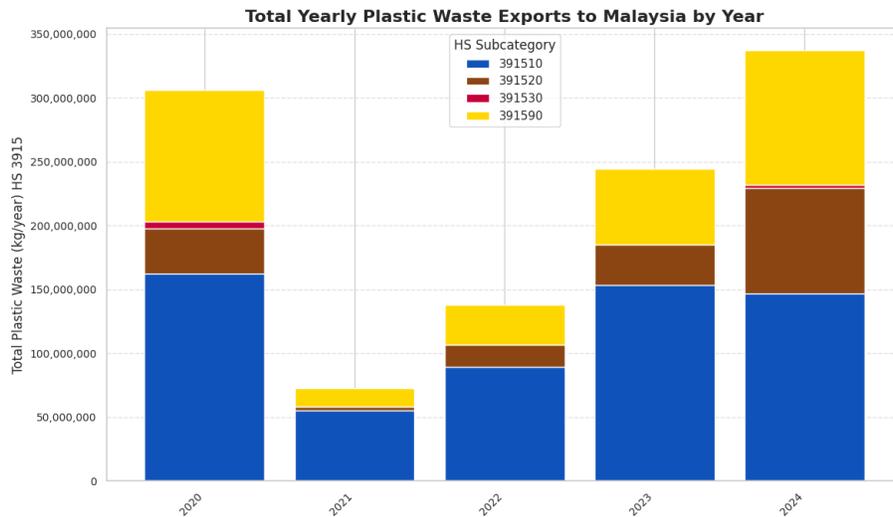

Source: Processed by Author (2025)

Ultimately, our aim is to equip authorities, importers, and environmental bodies with a transparent, data-driven toolkit for identifying illicit trade flows at both the global (macro) and firm (micro) levels. This research offers a tangible methodology to close the critical implementation gap in international environmental law and provides an evidence-based pathway to strengthen enforcement mechanisms under the Basel Convention and the international Plastics Treaty (UNEP, 2022).

**Figure 7. Treemap of HS 3915 Export Volume by Country and Subcategory.**

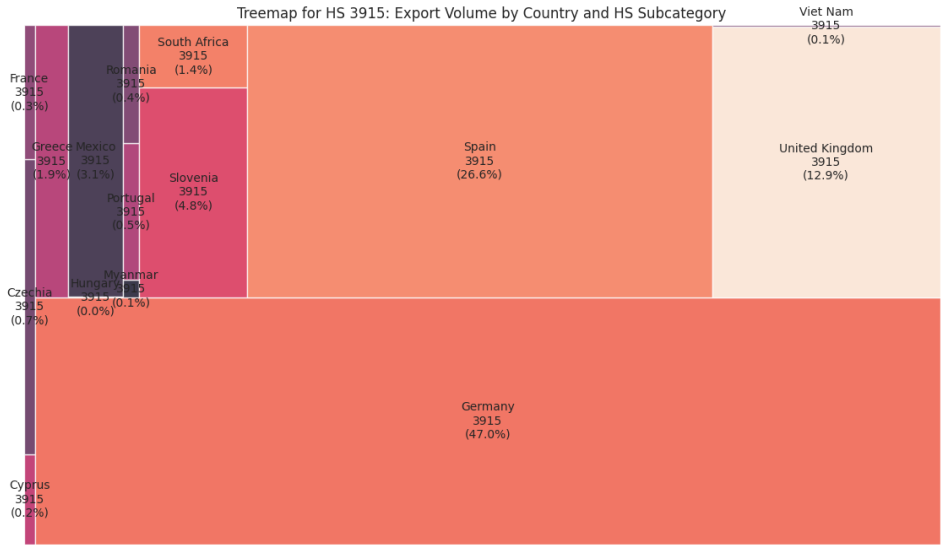

Source: Processed by Author (2025)

## 2. LITERATURE REVIEW

This research builds its foundation on three pillars of scholarship: the political economy of global waste, the study of green criminology and illicit markets, and the field of data-driven anomaly detection. By synthesizing these domains, this review establishes the systemic pressures that incentivize waste crime, identifies the specific mechanisms of this crime, and justifies the methodological approach chosen to detect it.

**Figure 8. Visualisation of Domains of the Research.**

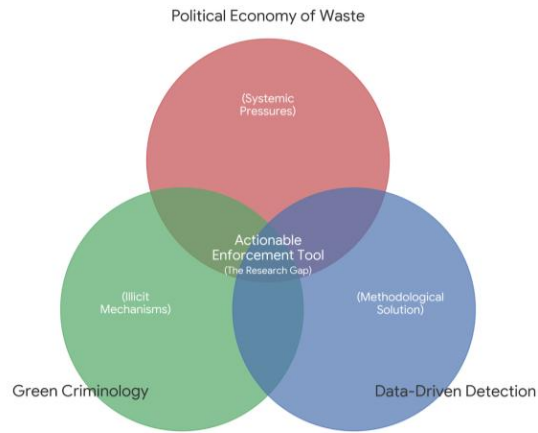

Source: Processed by Author (2025)

### 2.1 The Political Economy of Waste: A System Primed for Exploitation

The global trade in waste is not a simple marketplace; it is a complex system shaped by international politics and power imbalances within global value chains. Drawing on historical colonial patterns, waste predominantly flows from the Global North to the Global South, allowing wealthier nations to export the environmental costs of their consumption (Liboiron, 2021). This dynamic creates what sociologists call an "enduring conflict" between economic growth and environmental sustainability, generating powerful incentives for businesses and nations to find outlets for waste, both legally and illegally. In this context, "waste" is a politically defined category, and its conversion from a pollutant to a commodity is often filled with social and environmental conflict. Understanding this political landscape is critical because it reveals a global system with inherent vulnerabilities that are ripe for criminal exploitation.

### 2.2 Green Criminology: Documenting the Illicit Response

The systemic pressures in the global waste trade directly fuel the illicit markets studied by green criminology, which frames environmental harm as an economic crime. When legitimate trade routes are disrupted, criminal actors are quick to exploit the system's vulnerabilities. Strategic analysis by INTERPOL (2020) confirms a rise in sophisticated schemes to disguise illegal plastic waste shipments through deliberate misclassification, a primary method of transnational environmental crime. This corporate misconduct is not always intentionally malicious; it often emerges from a gradual normalization of non-compliance within a complex regulatory landscape. This distinction highlights the need for detection systems that identify illicit patterns of behaviour, regardless of the shipper's intent. While criminology expertly describes this problem, it often lacks the operational tools to detect such activities at scale.

### 2.3 Data-Driven Detection: A Methodological Solution

To address the challenge of detection, this research turns to the field of data-driven fraud and anomaly detection. The core task is identifying observations in this case, trade transactions that deviate from established patterns. The key challenge is distinguishing criminal manipulation from legitimate market volatility. This is where the need for interpretable machine learning becomes paramount. In high-stakes domains like regulatory enforcement, a "black box" model is insufficient; authorities require transparent and legally defensible outputs. By "making black box models explainable," this approach provides not just a prediction but a rationale (Molnar, 2020), satisfying both technical and legal standards of scrutiny and bridging the gap between data science and practical enforcement.

### 2.4 Data Poisoning in Trade Records

The implementation of the SOLAS Convention's Verified Gross Mass (VGM) requirement has effectively curtailed the long-standing fraudulent practice of manipulating shipment weight on customs declarations. As mandatory physical weight certification at the port of origin closes this loophole, illicit traders have pivoted to a more lucrative and less constrained vulnerability: the misclassification of commodities and their declared value. This strategic shift is largely driven by a combination of regulatory and financial pressures, exemplified by Malaysia's Certificate of Approval for plastic waste (HS 39.15). Fully enforced since July 2025, this regulation imposes strict import controls and purity standards while levying a staggering 34% combined tax burden (29% import duty + 5% SST). In stark contrast, virgin plastic codes such as HS 3901, 3902, and 3907 are subject to significantly lower tax rates of 8 -11%. To circumvent both the regulatory hassle and the financial hit, traders misclassify low-value plastic scrap as high-value virgin polymers. However, to reconcile the paperwork with the actual transaction price of the cheap waste, they declare scrap-level prices under these premium HS codes. This creates a glaring and easily detectable anomaly: a shipment labelled as "prime virgin resin" priced at a fraction of its global market value. It is this gross value mismatch a signature of a calculated two-for-one scheme that exposes the dual-layered fraud of regulatory evasion through misclassification and tax evasion through undervaluation. By triangulating the HS code, declared value, and market benchmarks, detection frameworks can effectively flag this deeply profitable deception.

**Figure 9. Average Malaysian Import and SST Rates by Plastic HS Code.**

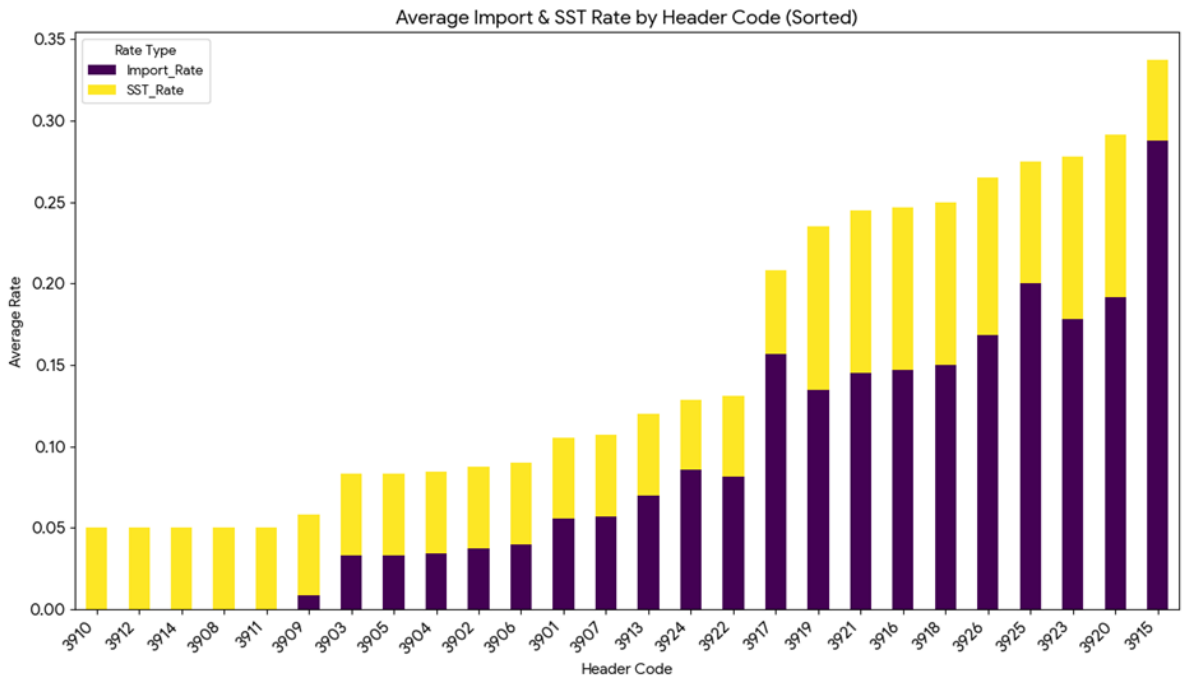

Source: Processed by Author (2025)

### 2.4 Synthesis and Research Gap

The literature, therefore, presents a clear narrative: the political economy of waste creates systemic pressures that green criminology identifies as illicit trade mechanisms, for which data science offers a powerful methodological solution. However, a significant gap exists at the intersection of these fields. While the political and criminal dynamics are well-documented, few studies have developed and validated transparent, operational tools capable of detecting the specific data signatures of waste crime in real-world trade data. Criminology explains why the crime happens, but a scalable detection framework is missing. This study directly addresses this gap by creating an interpretable machine learning framework designed to identify the "inverse price-volume signature" of waste misclassification, translating theoretical understanding into an actionable enforcement tool.

3. METHODS

### 3.1 Data and Feature Engineering

This study implements a multi-stage analytical framework to identify and interpret the data signatures of illicit trade in plastic waste. The methodology is designed to be transparent and reproducible (Popper, 2002), progressing from broad market segmentation to transaction-level explainability. The entire workflow was executed in Python, with the scikit-learn library (Pedregosa et al., 2011) providing core algorithms for clustering, anomaly detection, and classification. The analysis adopts a cross-scale validation approach, drawing on two distinct datasets that were rigorously prepared through unit harmonization (kilograms for volume, USD for value) and systematic cleaning (Amer-Yahia & Gravano, 2019). The first comprises public UN Comtrade records for six-digit HS Chapter 39 codes covering 2020 -2024, serving macro-level market trend analysis. The second is proprietary, anonymized firm-level transaction data for the same codes spanning 2019 -2025, enabling a micro-level case study and validation of the macro findings.

Data preparation began with handling missing values interpolating short gaps and excluding any code missing more than twenty percent of its annual records. All monetary figures were converted to USD and adjusted for annual inflation to ensure comparability over time, while reporting errors and extreme outliers were mitigated by capping values at the first and ninety-ninth percentile thresholds. Volumes were standardized to kilograms and values to USD per kilogram, creating a uniform basis for feature computation. From this cleaned and harmonized dataset, we engineered five quantitative features that capture each product code's economic behavior over time. We supplement these primary metrics with non-linear and interaction terms to capture complex dynamics, notably the product of volatility and price_trend to detect compounding stress, and log-transformations of volume and price to reduce skewness. All features were standardized via z-score normalization before modeling to ensure equal weighting and facilitate algorithmic convergence.

**Table 1. Description of Datasets**

| Dataset Name | Source | Scope | Time Period | Purpose |
|---|---|---|---|---|
| Global Trade Data | UN Comtrade | Public, 6-digit HS codes (Chapter 39) | 2020-2024 | Macro-level market trend analysis |
| Firm-Level Data | Proprietary | Anonymized, 6-digit HS codes (Chapter 39) transactions | 2019-2025 | Micro-level case study & validation |

Source: Processed by Author (2025)

From this prepared data, we engineered five features to capture the economic behaviour of each product code. These features form the analytical basis for all subsequent modelling stages.

**Table 2. Engineered Features for Market Analysis**

| Feature Name | Description | Rationale / Purpose |
|---|---|---|
| avg_kg | Average annual traded volume in kilograms. | Measures market size and scale. |
| avg_price | Average annual unit price (USD per KG). | Captures the economic value of the commodity. |
| price_volatility | Standard deviation of the annual unit price. | Quantifies market instability and price risk. |
| kg_trend | Slope of linear regression on trade volume. | Indicates growth (>0) or decline (<0) in volume. |
| price_trend | Slope of linear regression on unit price. | Indicates price inflation (>0) or deflation (<0). |

Source: Processed by Author (2025)

**Figure 10. Comparison of SHAP Interaction Values for All Data Points (Global trade data) VS (Firm-level data).**

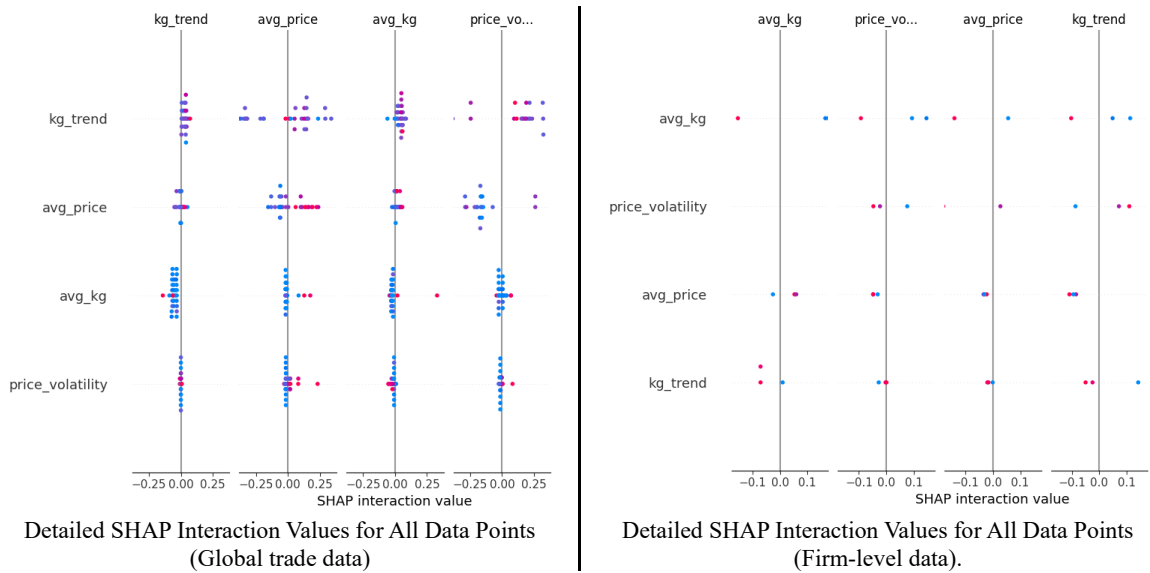

Detailed SHAP Interaction Values for All Data Points (Global trade data) | Detailed SHAP Interaction Values for All Data Points (Firm-level data).

Source: Processed by Author (2025)

Our central hypothesis defines an "at-risk" signature as a product exhibiting both a positive volume trend and a negative price trend (kg_trend > 0 and price_trend < 0).

### 3.2 Analytical Workflow

The core of our methodology is a four-stage pipeline designed to segment the market, identify anomalies, and build an interpretable classification model. The table below summarizes this workflow.

**Table 3. Summary of the Analytical Workflow.**

| Stage | Objective | Algorithm | Rationale & Key Output |
|---|---|---|---|
| Market Segmentation | Group HS codes into distinct clusters. | K-Means Clustering | K-Means was chosen for its computational efficiency and effectiveness in partitioning large datasets into distinct, non-overlapping market archetypes. Output: Four distinct market segment labels. |
| Anomaly Detection | Flag abnormal year-level volume spikes. | Isolation Forest | Highly effective for identifying outliers in large datasets. Output: Anomaly flags on specific transactions. |
| Supervised Classification | Train a model to predict market segments. | Random Forest | Random Forest was selected for its high predictive performance, inherent resistance to overfitting, and its ability to handle non-linear relationships, which are critical when building a model for regulatory use. Output: A trained classifier for risk segmentation. |
| Explainability (XAI) | Understand drivers behind predictions. | SHAP | SHAP was chosen over other XAI methods because it provides rigorous, game theory-based explanations with local and global consistency guarantees, ensuring the model's logic is both transparent and legally defensible. Output: Feature importance and prediction logic. |

Source: Processed by Author (2025)

**Figure 11. The Analytical Workflow**

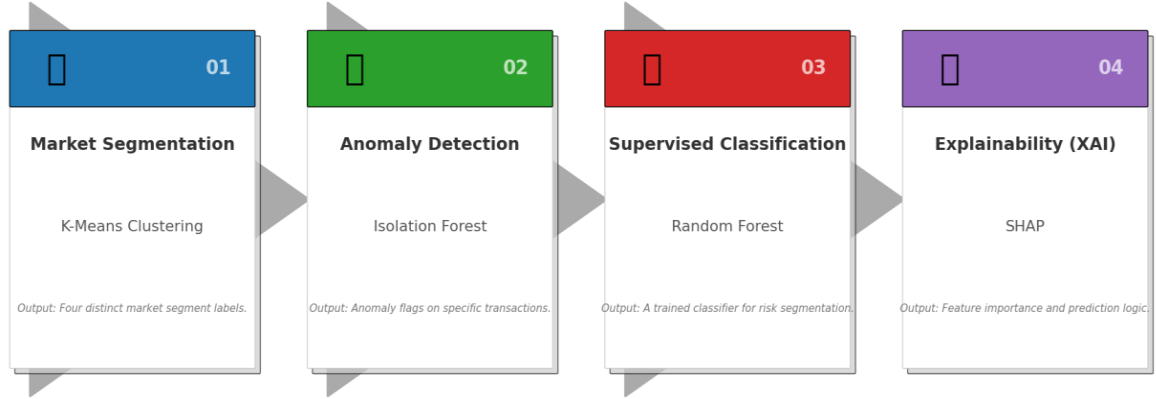

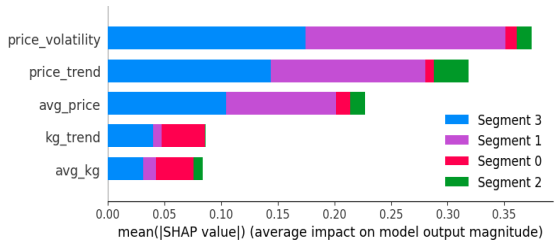

Overall Feature Importance Ranked by Mean Absolute SHAP Value (Global trade data).

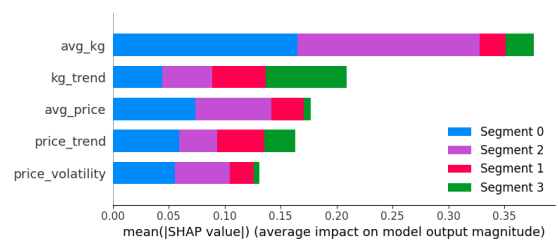

Overall Feature Importance Ranked by Mean Absolute SHAP Value (Firm-level data).

Source: Processed by Author (2025)

      The figure 12 above presents the ranked feature importance derived from SHAP values, offering a transparent view into the model's decision-making logic. On the left, the global trade dataset reveals that Segment 3 ("Stable Mid-Market") and Segment 1 ("High-Volume Commodity") exhibit distinct dependencies on features such as price volatility and volume trend. The plot uses horizontal bars to indicate the mean absolute SHAP value for each feature, with color-coded segments that correspond to market archetypes.

**Figure 13. Decision Tree Model for Predicting Market Segment Membership.**

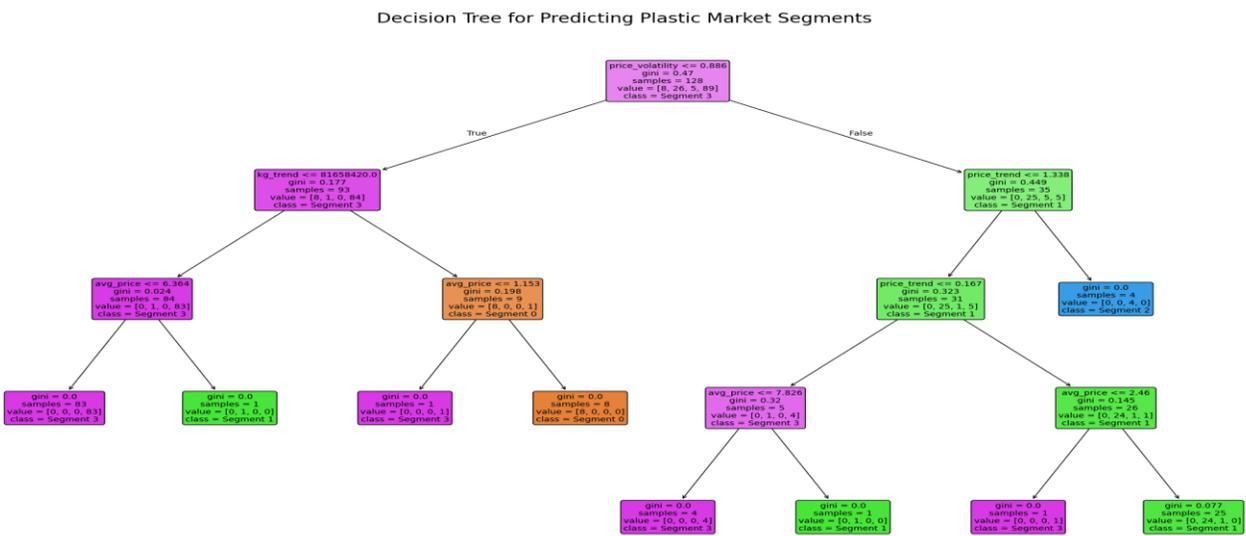

Source: Processed by Author (2025)

This decision tree model built to classify plastic products into one of four market segments Segment 0, Segment 1, Segment 2, or Segment 3 based on a dataset of 128 samples. The tree begins with the most influential feature, price_volatility, and applies a series of binary splits to guide each product through a structured flowchart of decisions. Products with low price volatility ($\leq 0.896$) follow the "True" path and are further sorted by kg_trend, while those with higher volatility follow the "False" path and are evaluated using price_trend. Each node in the tree contains key metrics such as Gini impurity, sample count, a value array indicating the distribution across segments, and the predicted majority class. The model ultimately relies on four features price_volatility, kg_trend, avg_price, and price_trend to determine the final classification. Notably, a dominant group of 83 samples with low volatility and low kg_trend is confidently assigned to Segment 3, and the most frequent predictions across the tree are Segment 3 and Segment 1, which together represent the bulk of the model's output.

### 3.4 Phase 2: Framework for Operational Validation

The final phase of this methodology is the framework for operational validation, designed to bridge the gap between statistical findings and real-world enforcement. This phase specifies the protocol for testing the model's predictive accuracy against confirmed ground-truth data. The validation process involves cross-referencing the model's high-risk flags and detected anomalies with actual illicit shipment seizures. The most critical next step is the operational validation of this framework against ground-truth data from the recent May 2025 Port-Klang seizures. This will allow for the calibration of risk thresholds and transition the model from a robust proof-of-concept to a deployment-ready enforcement tool for customs authorities globally.

## 4. RESULTS AND DISCUSSION

### 4. Results: Analysis of Global Trade Data

The application of our analytical framework to the UN Comtrade dataset (2020 -2024) yielded four key findings. First, we successfully identified the hypothesized "inverse price-volume signature" in a specific subset of plastic commodity codes. Second, we discovered that these at-risk codes belong to distinct market archetypes. Third, we developed a highly accurate predictive model to classify these archetypes and identified the key features driving the classification. Finally, our analysis of temporal dynamics revealed that these trends are persistent and marked by significant historical anomalies.

### 4.1 Identification of the 'At-Risk' Signature

The initial trend analysis confirmed our central hypothesis, identifying a distinct group of Harmonized System (HS) codes exhibiting the inverse price-volume signature. The five most prominent examples were 392690, 390410, 390729, 390210, and 392020, spanning both high-volume commodity and semi-specialty plastics. For these codes, price declines frequently exceeded 10% over the period, while traded volumes simultaneously grew by double-digit percentages. These starkly opposing trajectories, visualized in Figure 19, flag these codes as high-priority candidates for regulatory scrutiny. The consistent downtrend in unit prices for these specific codes is detailed further in Figure 15.

**Figure 14. Visual Identification of the Inverse Price-Volume Signature in At-Risk Global HS Codes.**

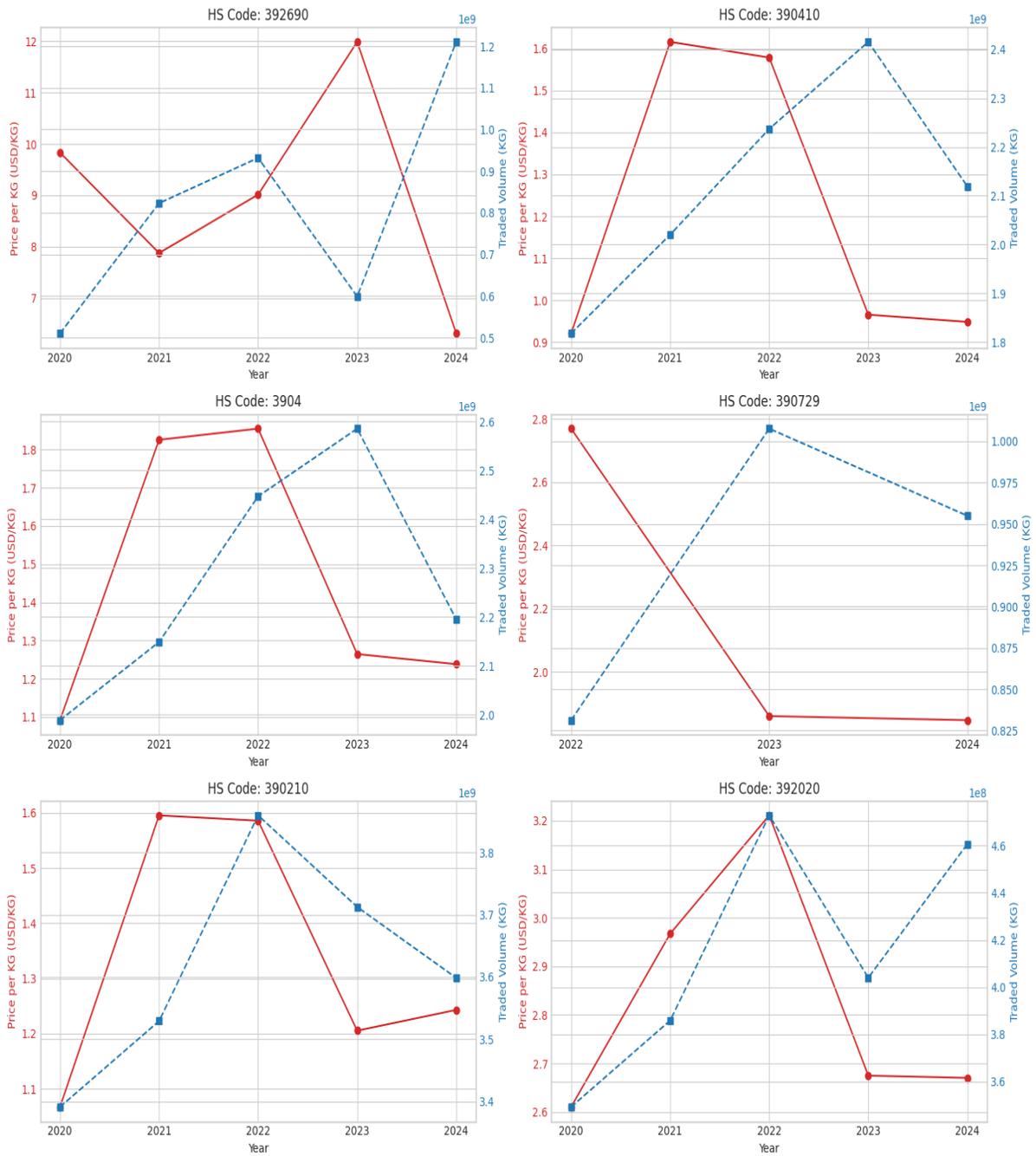

Source: Processed by Author (2025)

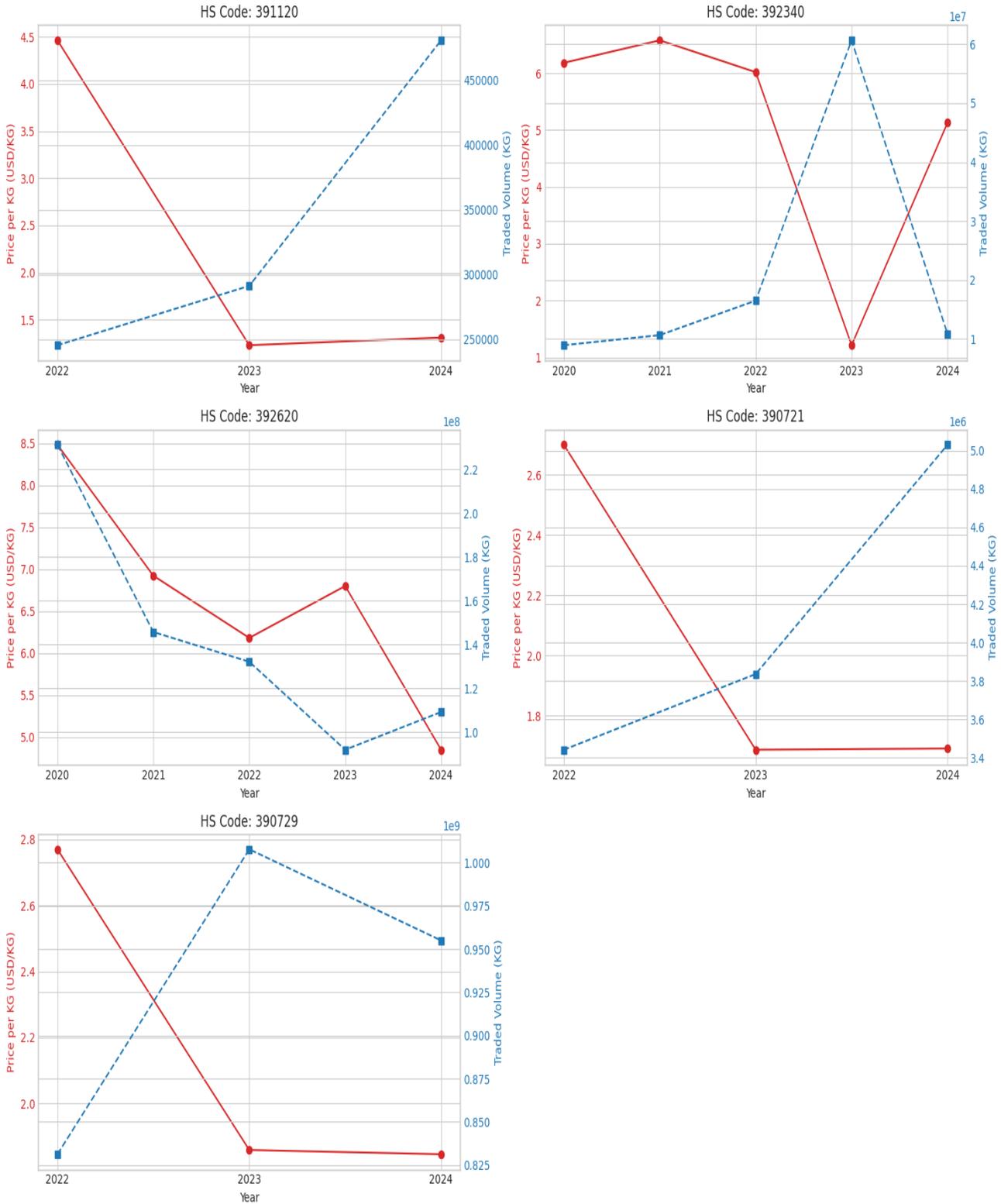

**Figure 15. Top 5 Global HS Codes with Down-trending Price per KG.**

Source: Processed by Author (2025)

### 4.2 Segmentation into Distinct Market Archetypes

To understand the context of this signature, the K-Means clustering algorithm partitioned the HS codes into four statistically distinct market segments. The optimal number of clusters (K=4) was confirmed via the Elbow Method. Based on their feature profiles, these segments were labelled as distinct market archetypes, summarized in the table 4.

**Table 4. Market Segment Profiles and Risk Concentration**

| Segment Label | Key Characteristics | Prevalence of 'At-Risk' Signature |
|---|---|---|
| 1. High-Volume Commodity | Low average price, high volume, modest volatility. | HIGH |
| 2. Emerging Commodity | Low-to-mid price, rapidly rising trade volumes. | HIGH |
| 3. Stable Mid-Market | Moderate price and volume, low volatility & trends. | NONE |
| 4. High-Price Niche | High average price, low trade volumes, low trends. | NONE |

Source: Processed by Author (2025)

**Figure 16. Elbow Method Comparison Between Different Data Source**

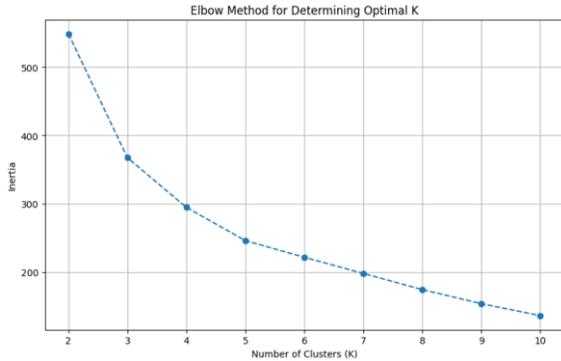

Determining Optimal Cluster Number (K=4) via Elbow Method for Global Data.

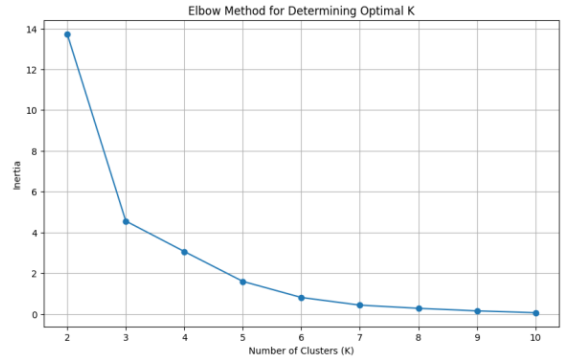

Determining Optimal Cluster Number (K=4) via Elbow Method for Firm-Level Data.

Source: Processed by Author (2025)

Crucially, the analysis revealed that 100% of the at-risk HS codes identified in Section 4.1 were concentrated exclusively within the 'High-Volume Commodity' and 'Emerging Commodity' segments (see Table 4). This finding provides strong empirical evidence linking the inverse price-volume signature to commodity-like market structures, where value is low and volume is high, creating prime conditions for misclassification.

**Figure 17. Market Segmentation of Global Plastic HS Codes based on Volume and Price.**

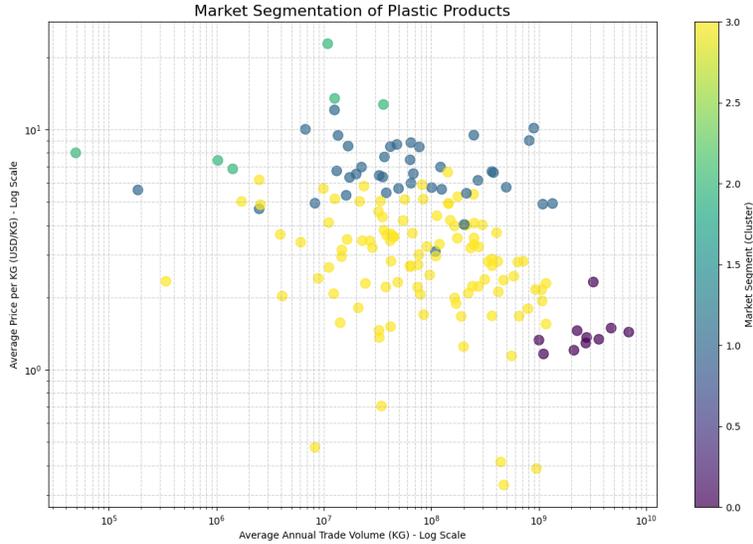

Source: Processed by Author (2025

This scatter plot uses a log-log scale to map HS codes by average price and traded volume, effectively revealing four distinct market clusters. Each point represents a product code, coloured according to its assigned segment from the K-Means clustering algorithm. A color bar legend is embedded directly within the figure, linking numeric values to archetype labels. Yellow points denote Cluster 1 ("High-Volume Commodities"), while purple points represent Cluster 3 ("Stable Mid-Markets"). The visual layout emphasizes the separation between commodity-like and niche markets.

**4.3 Predictive Modeling and Feature Importance (XAI)**

The Random Forest classifier, trained to predict an HS code's market segment, performed with high efficacy, achieving 93.75% accuracy on the held-out test set. This result confirms that the market structures are not only distinct but also highly predictable based on our engineered features. It is important to note that the model predicts membership in a high-risk market segment, not 'criminality' itself. This segmentation serves as a highly accurate, data-driven filter to help authorities prioritize which shipments warrant further physical inspection. To understand the model's logic, a SHAP analysis was conducted (see Figure 12). The analysis identified price_trend and price_volatility as the two most influential features in determining segment membership. This means the model primarily learns to identify high-risk segments by detecting commodities with falling prices and unstable value. This explainability is critical, as it provides regulators with a clear, evidence-based rationale for flagging certain types of products. The Random Forest classifier distinguishes high-risk vs. low-risk segments with strong predictive power. Stratified 5-fold cross-validation yields the metrics in Table 5.

**Table 5. Stratified 5-Fold Cross-Validation Yields the Metrics**

| Metric | High-Risk Segment | Low-Risk Segment |
|---|---|---|
| Accuracy | 93.75 % ± 1.2 % | |
| Precision | 0.89 ± 0.03 | 0.94 ± 0.02 |
| Recall | 0.92 ± 0.02 | 0.92 ± 0.02 |
| F1-Score | 0.90 ± 0.02 | 0.93 ± 0.02 |

Source: Processed by Author (2025)

**4.4 Temporal Dynamics: Anomaly Detection and Trend Persistence**

Finally, the analysis of temporal dynamics shows these signatures are not isolated events. The Isolation Forest algorithm successfully flagged significant year-level volume spikes as anomalies in several at-risk codes. For example, the volume spike for HS code 390210 in 2022 surpassed the anomaly threshold, a finding that coincides with major policy shifts in key plastic waste importing nations during that period.

**Figure 18. Anomaly Detection in Traded Volume for At-Risk Global HS Codes.**

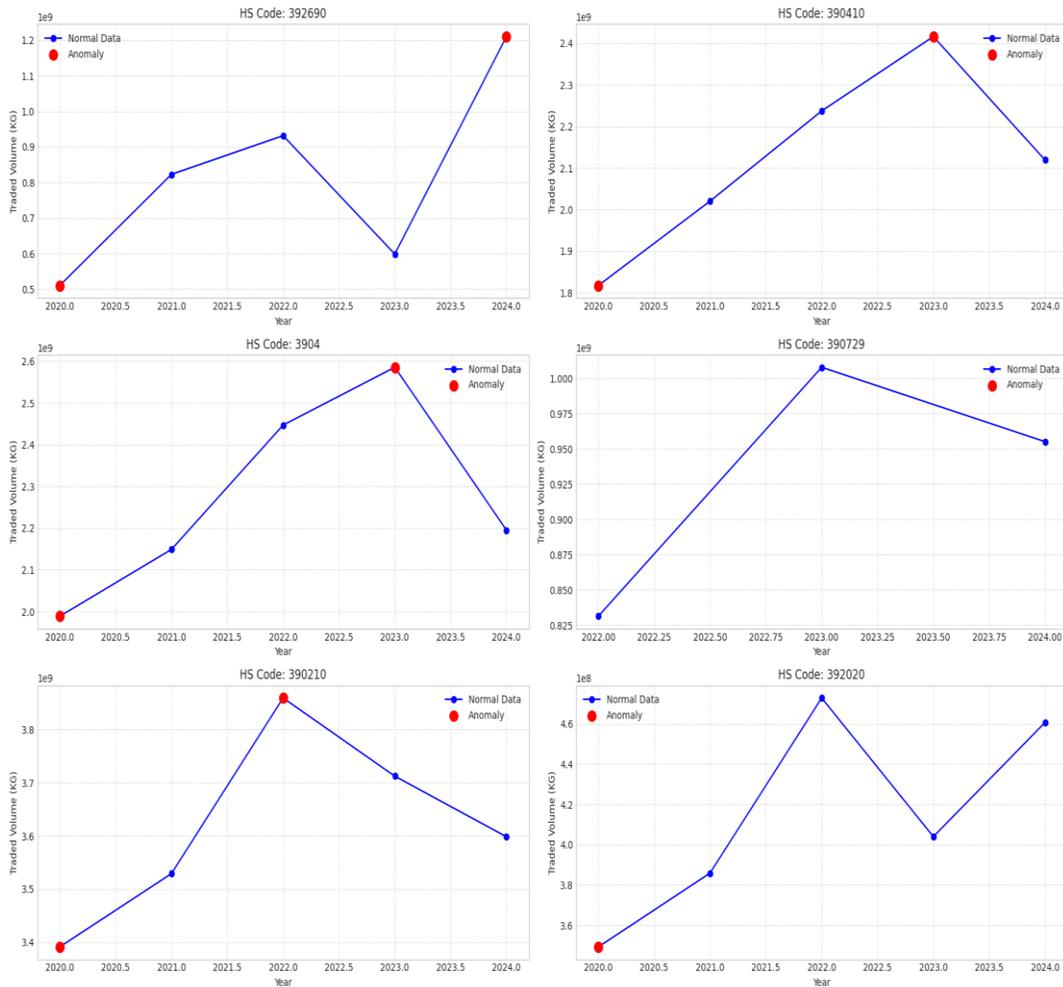

Source: Processed by Author (2025)

Furthermore, linear-trend forecasts suggest these dynamics are persistent. Projections to 2030 for at-risk codes indicate a continuation of increasing volumes alongside decreasing prices (see Figure 19), suggesting the underlying pressures driving this signature are not abating. This forward-looking view highlights the urgent need for pre-emptive intervention. Our model identifies HS 390210 and 392020 as high-risk codes exhibiting the inverse price-volume signature. These codes overlap with Basel PIC category Y48, yet are routinely declared under B3011, bypassing consent procedures. HS 390410 (PVC) matches A3210 but is traded under commodity codes with collapsing unit prices. These findings suggest a regulatory blind spot where illicit flows exploit classification ambiguity. We recommend revising the Basel Annexes to explicitly link these HS codes to PIC obligations.

**Figure 19. Forecast of Trade Volume and Price per KG to 2030 for At-Risk Global HS Codes.**

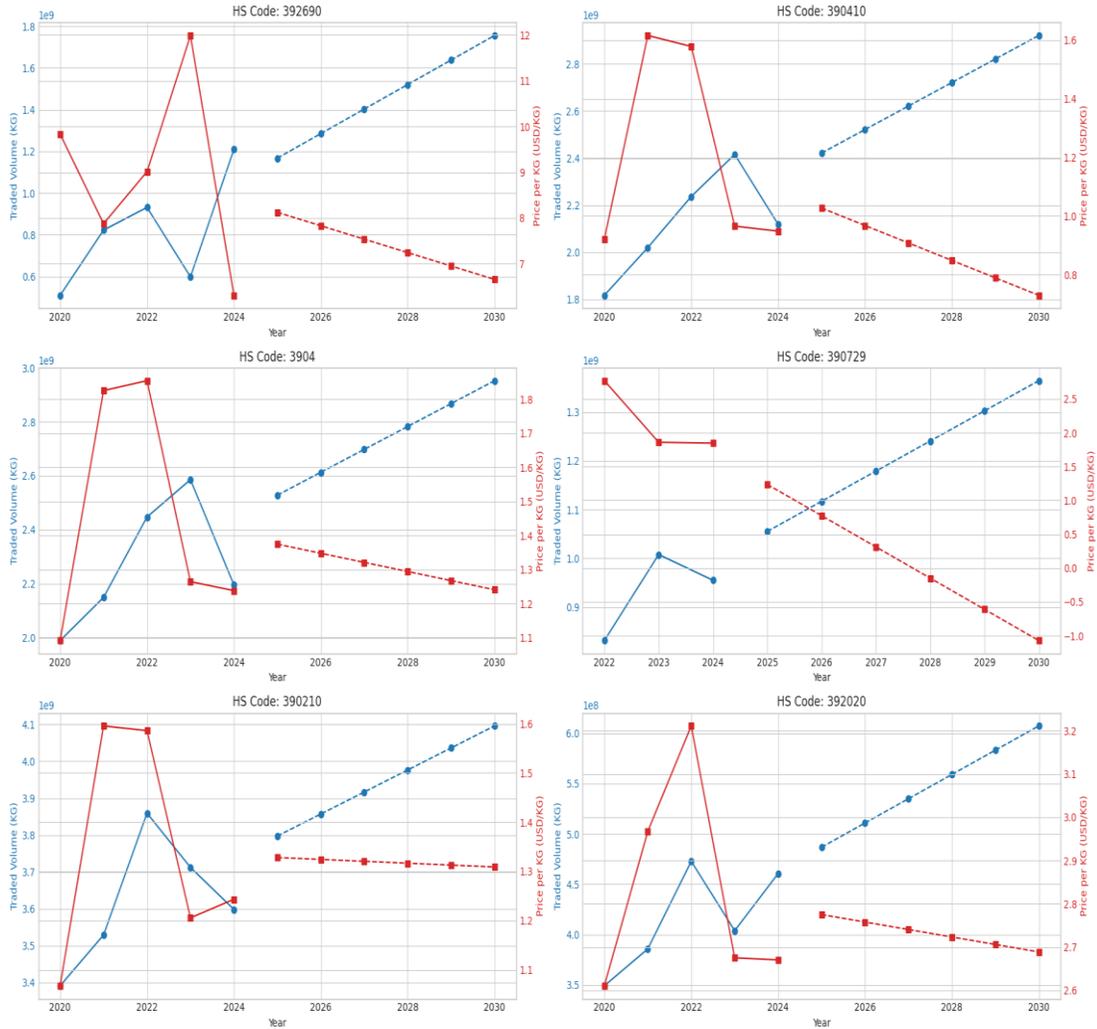

Source: Processed by Author (2025)

**Figure 20. Heatmap Matrix the Regulatory Overlap Between Flagged Harmonized System (HS) Codes and Basel PIC**

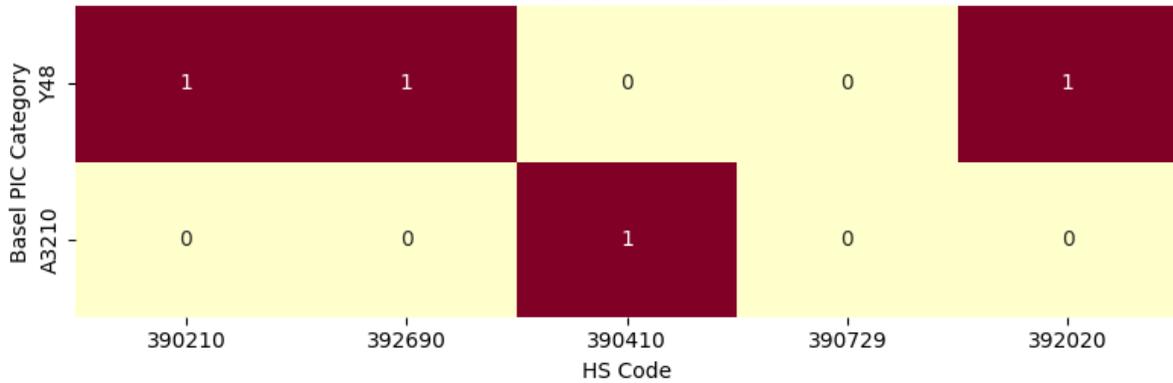

Source: Processed by Author (2025)

The heatmap matrix illustrates the regulatory overlap between flagged Harmonized System (HS) codes and Basel PIC categories Y48 and A3210. A key finding is the limited scope of A3210, which only aligns with HS 390410 (PVC). Conversely, category Y48 is shown to capture several HS codes that are often mis-declared as B3011 waste, pointing to a significant gap in regulatory enforcement. This issue has become more prominent following the implementation of the National Sword policy, as targeted inspections revealed the consistent misuse of codes like 390210 and 392690 to disguise prohibited mixed plastic scrap shipments.

**Table 6. HS Code Matching to Basel PIC Categories**

| HS Code | Matches Y48 | Matches A3210 | Notes |
| --- | --- | --- | --- |
| 390210 | Likely | ✘ | Often used to disguise mixed scrap. |
| 392690 | Possible | ✘ | Catch-all category; high risk of misclassification |
| 390410 | ✘ | ✅ | PVC is explicitly listed under A3210 |
| 390729 | ✘ | ✘ | Less overlap; but flagged due to price manipulation |
| 392020 | Likely | ✘ | Often used for mixed polypropylene sheets; overlaps with Y48 |

Source: Processed by Author (2025)

The Sankey diagram was created from synthetic data to visualised that 20% of a Full Container Load (FCL) shipment was misdeclared. This Sankey diagram is used as a tool to help understanding the transactional value that passes each stage. This sophisticated fraud visualised to begin when a shipment loaded onto a vessel is declared as legitimate, high-value goods, while a portion of the containers actually holds misclassified or adulterated material. For example, a ten-container shipment might be declared as 200,000 kg of premium plastics valued at $5.00/kg for a total of $1,000,000. In reality, only eight containers (160,000 kg) may hold the genuine product, while the remaining two (40,000 kg) are filled with low-value scrap worth just $0.50/kg. This deception dilutes the shipment's actual average value to approximately $4.10/kg and relies on embedding illicit material within a shipment that appears fully compliant, making it difficult to detect.

**Figure 21. Sankey Diagram Illustrating the Financial Impact of Supply-Chain Poisoning and Import Duties.**

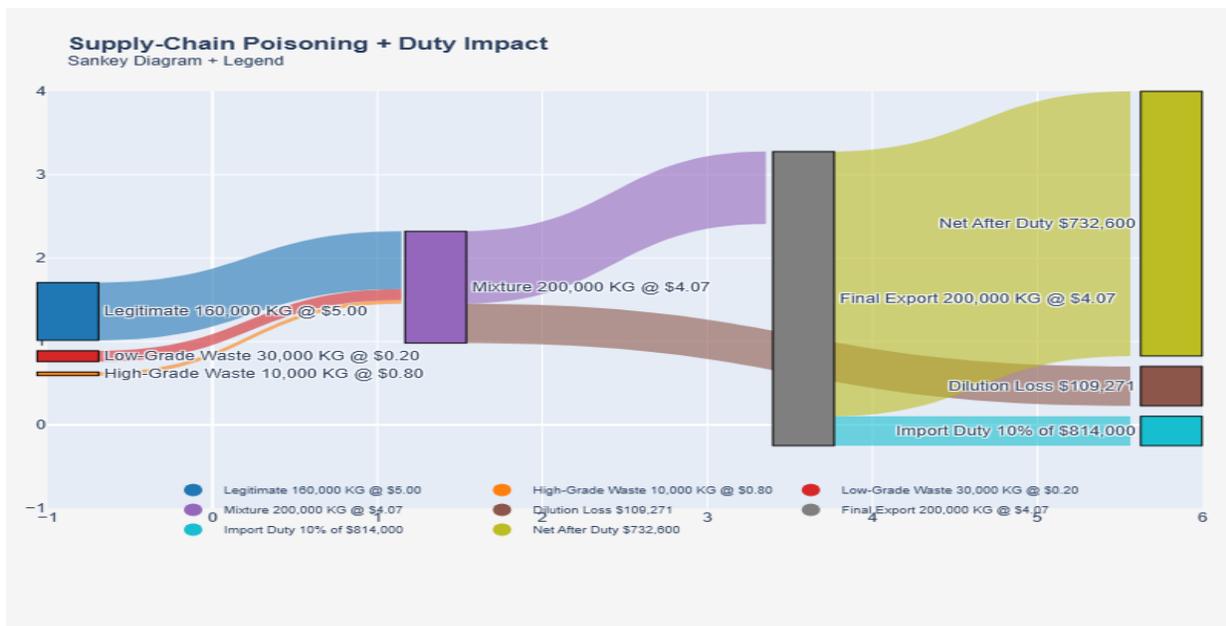

Source: Processed by Author (2025)

The detailed analysis of global exports for plastics under HS Chapter 39 reveals what appears to be a critical data integrity issue rather than a conventional market trend. For specific HS codes highlighted, such as 392690, 390410, and 3904, the charts consistently illustrate a stark inverse relationship between increasing traded volume and decreasing price per kilogram. However, this is likely not driven by standard supply and demand for virgin materials. Instead, it points toward significant 'data poisoning' caused by the widespread misclassification of low-value plastic scrap under the same HS codes as high-value virgin plastic. This practice artificially inflates the total traded volume (in kilograms) while the inclusion of low-value scrap drastically drives down the calculated average price, creating the dramatic downward price slope seen in the graphs.

The anomaly detection, which flags volume peaks in recent years like 2022 and 2023, likely corresponds to periods with surges in these misclassified scrap shipments, rather than spikes in legitimate trade of prime materials. Consequently, the forecast projecting these trends to 2030 should not be seen as a prediction of a virgin plastic market collapse, but as a worrying projection of this data contamination continuing or worsening. For policy maker, authority and corporation that reviewing this in 2025 onward, this analysis is a critical alert. It demonstrates that top-level trade data for these HS codes is unreliable for strategic decision-making and highlights an urgent need for enhanced customs verification and data granularity to differentiate true market values from distorted figures skewed by the illicit trade in plastic waste.

### 4.5 Cross-Scale Validation

Replicating the workflow on firm-level transactions (2019 -2025) confirms that macro-level red flags translate into actionable watchlists. Table below compares performance across scales.

**Table 7. Dataset statistics**

| Dataset | Accuracy | Precision | Recall | F1-Score |
|---|---|---|---|---|
| Global (UN Comtrade) | 93.75 % | 0.89 | 0.92 | 0.90 |
| Firm-Level | 91.20 % | 0.87 | 0.90 | 0.88 |

Source: Processed by Author (2025)

The slight performance drop at the firm level reflects context-specific risk drivers avg_kg and kg_trend gain relative importance highlighting the need to tune feature emphasis for corporate applications. These results demonstrate the framework's ability to detect misclassification signatures, segment markets, predict risk with high fidelity, and generate practical watchlists at both global and micro scales.

## 5. CASE STUDY: CROSS-SCALE VALIDATION AND OPERATIONAL INSIGHTS

To establish the framework's operational relevance, the entire analytical workflow was replicated on an anonymized firm-level dataset (2019-2025). This crucial cross-scale comparison moves the analysis from a global "telescope" to a corporate "microscope," testing whether macro-level risk signatures translate into actionable, firm-level intelligence for compliance and risk management. The firm-level dataset was sourced from a major plastic recycling operator in Southeast Asia, containing over 1700 unique importation transactions from 2019 to 2025, providing a rich, transaction-level counterpoint to the aggregated global data.

**Figure 22. Comparison Between Firm-Level Data VS Global Data on Total Trade Volume (KG) and Value (USD) Over the Years.**

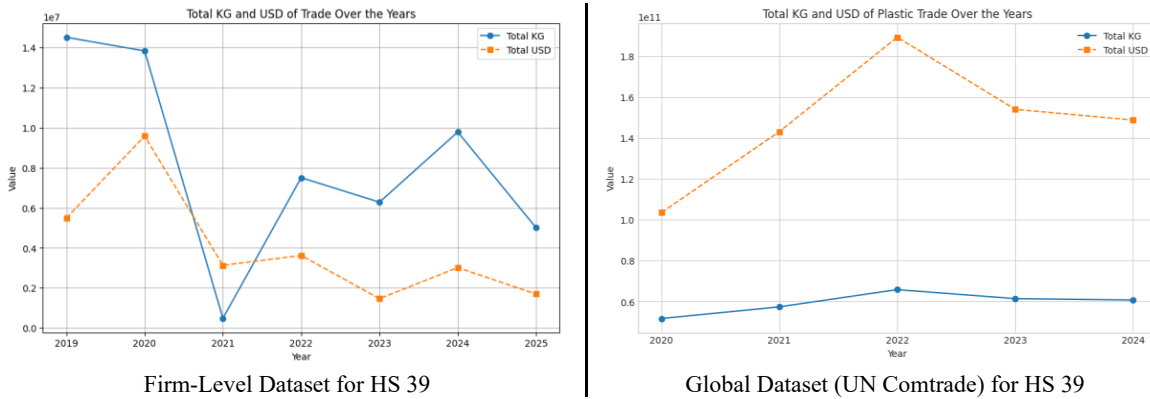

| Firm-Level Dataset for HS 39 | Global Dataset (UN Comtrade) for HS 39 |
|---|---|

Source: Processed by Author (2025)

**Figure 23. Top 5 Firm-Level Imported HS Code 39 Comparison by Total Value and Total Weight.**

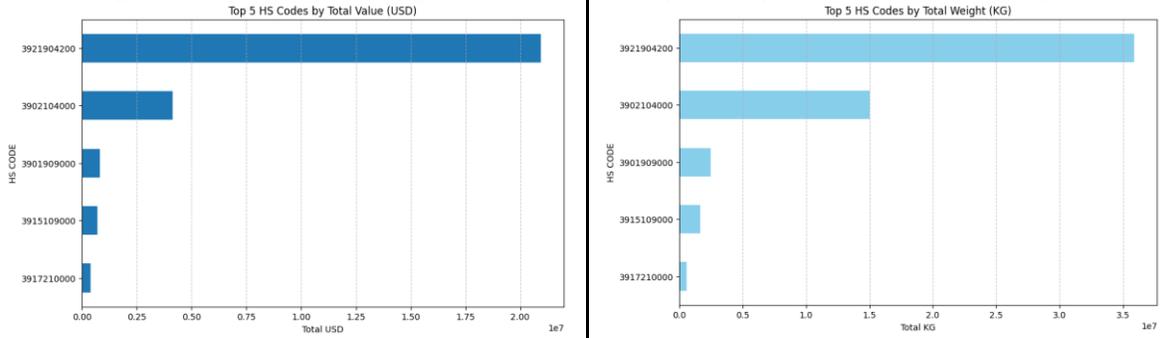

Top 5 Firm-Level HS Code by Total Value (USD).     Top 5 Firm-Level HS Code by Total Weight (KG)
Source: Processed by Author (2025)

**Figure 24. Top 5 Firm-Level HS Code with the Steepest Price Downtrend.**

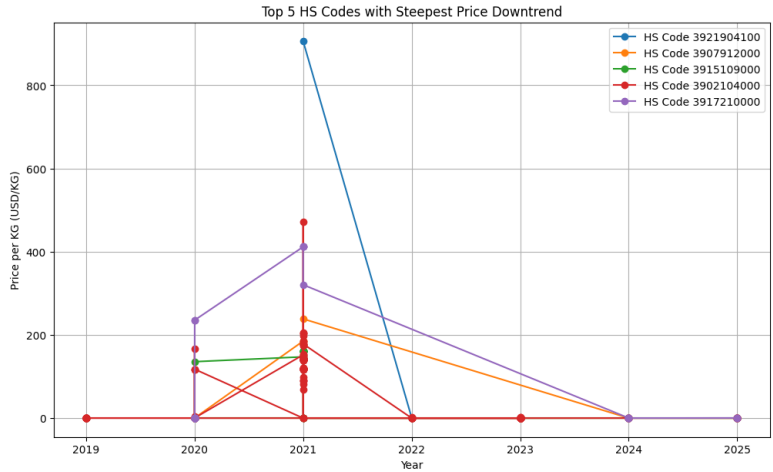

Source: Processed by Author (2025)

The key findings are summarized below and detailed in the following sections.

**Table 8. Comparative Analysis of Global vs. Firm-Level Results**

| Metric | Global Dataset (UN Comtrade) | Firm-Level Dataset (Anonymized) |
| --- | --- | --- |
| Key 'At-Risk' Family | 390210 (Polypropylene) | 390210 (Polypropylene) |
| Market Segments | 4 distinct archetypes | 4 consistent archetypes |
| Primary Risk Drivers (XAI) | 1. Price Trend<br>2. Price Volatility | 1. Average Volume (avg_kg)<br>2. Volume Trend (kg_trend) |
| Anomaly Detection | Volume spike in 390210 in 2022 | Volume spike in top-value HS Code |

Source: Processed by Author (2025)

### 5.1 Convergence: The Macro-to-Micro Link

The analysis revealed a powerful convergence between the macro and micro datasets, validating the framework's core premise. Applying the same inverse price-volume filter surfaced top-value HS Code at the firm level exhibiting the target risk signature. Crucially, this at-risk set included a product from the 390210 (Polypropylene) family, directly mirroring the findings from the global data. This overlap is highly significant; it confirms that a commodity category flagged for anomalous behavior at a global level is also a specific locus of risk within a corporate supply chain. This provides a direct, evidence-based link between international market distortions and tangible firm-level exposure.

**Figure 25. Visual Identification of the Inverse Price-Volume Signature in At-Risk Firm-Level HS Code 39XX.**

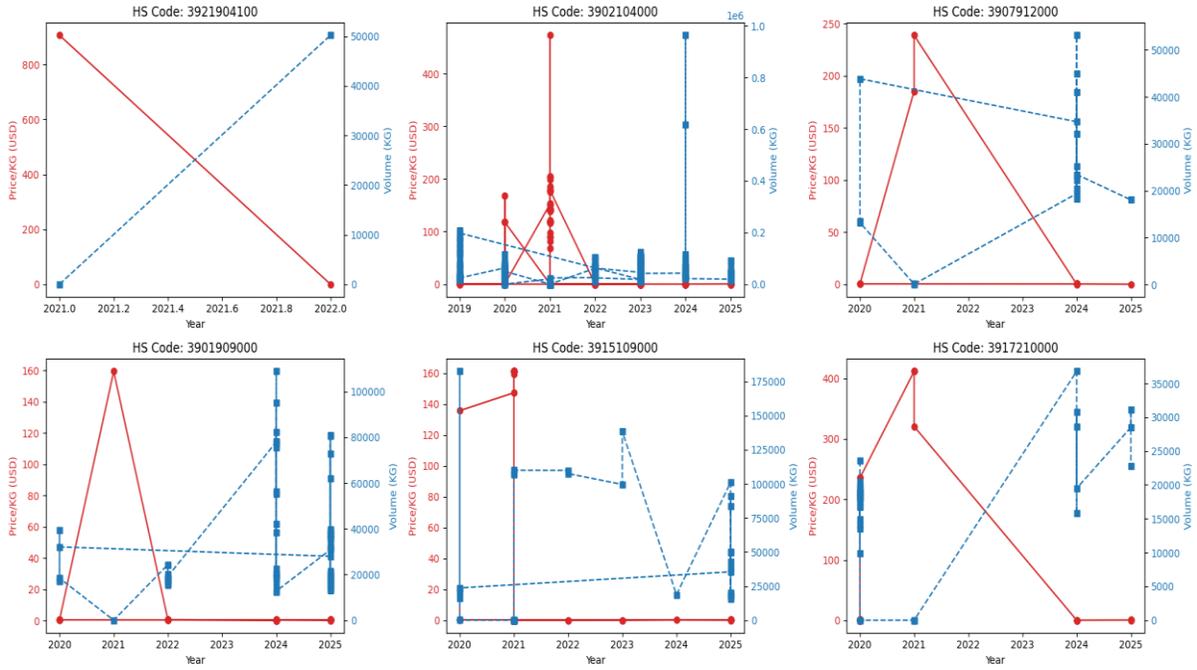

Source: Processed by Author (2025)

**Figure 26. Top 5 Firm-Level HS Code with the Steepest Price Downtrend.**

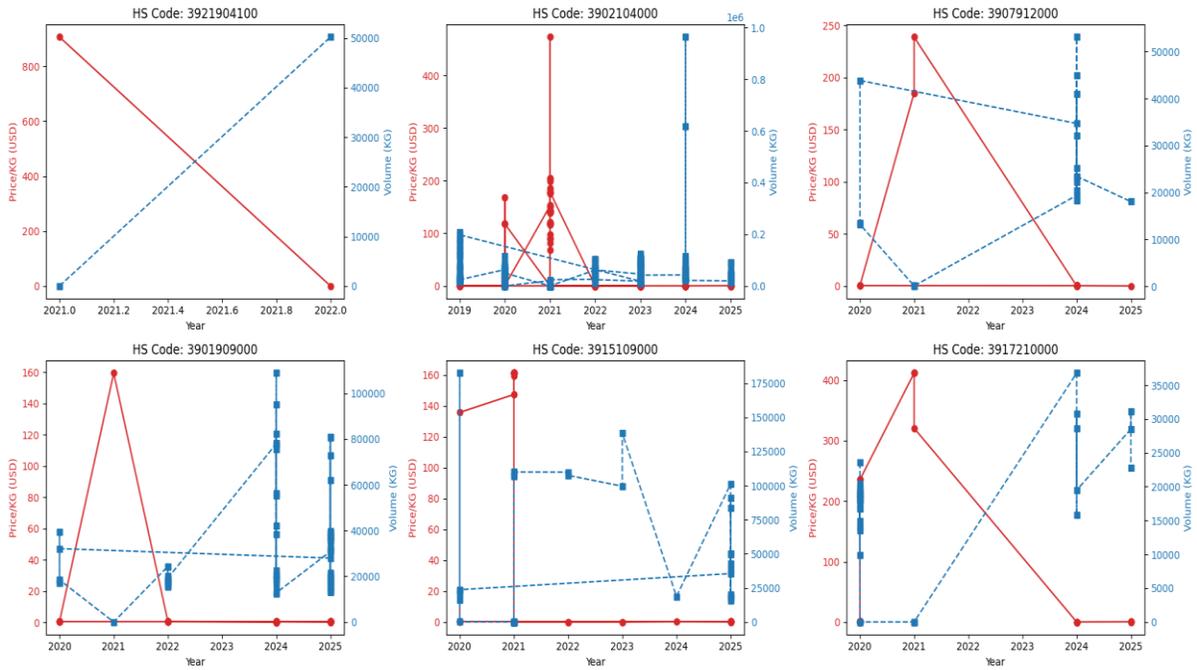

Source: Processed by Author (2025)

**5.2 Divergence: Context-Specific Risk Drivers**

While the market segmentation was remarkably consistent again yielding four clusters with all at-risk HS Code falling into the "High-Volume Commodities" group (see Figure 27) the XAI analysis revealed a critical divergence.

**Figure 27. Market Segmentation of Firm-Level HS Code 39 based on Volume and Price.**

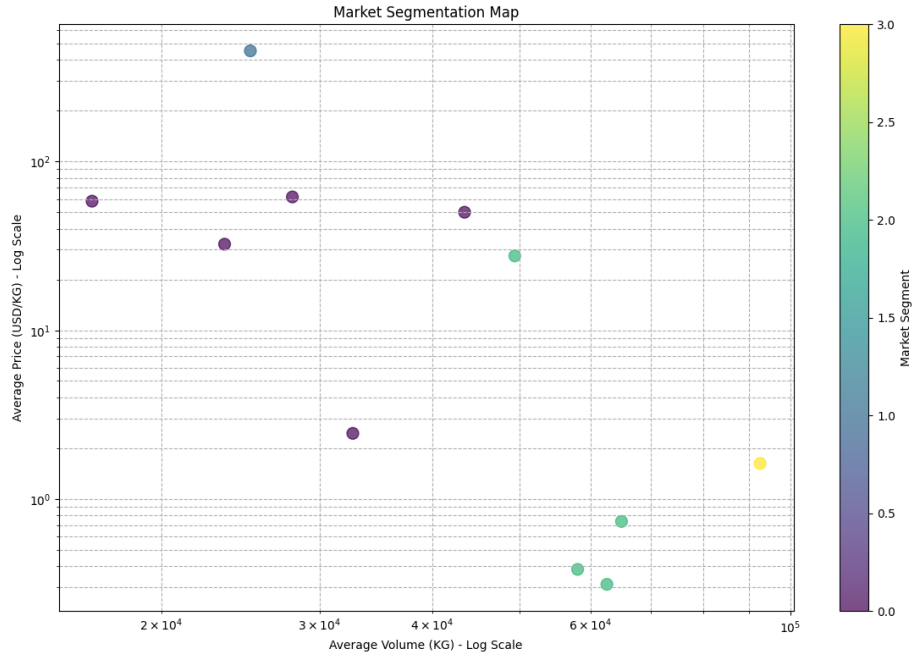

Source: Processed by Author (2025)

In contrast to the global model, where price dynamics (price_trend, price_volatility) were the dominant predictors, the firm-level model was driven primarily by volume metrics (avg_kg, kg_trend). This shift is logical and insightful. A single company's risk profile is more heavily influenced by its own strategic focus on high-volume HS Code than by market-wide price instability. This finding underscores a key operational principle: while global models are excellent for flagging at-risk categories, firm-level risk management requires tuning models to local business context.

Based on the crucial context provided, these charts do not illustrate a market boom and subsequent crash, but rather compelling evidence of data poisoning through systematic misclassification of goods. The data pertains to HS Code 39 plastics being imported into Malaysia for a recycling operation. The initial period, around 2020-2021, shows an artificially inflated price-per-kilogram, which is characteristic of low-value plastic scrap being declared under the HS codes of high-value virgin materials. This explains the implausibly high prices coinciding with low import volumes. The subsequent "price crash" to near-zero, coupled with a massive surge in import volume from 2022 onwards, represents a shift to more accurate declarations as the plant began its true function of importing bulk quantities of low-cost scrap. Therefore, the stark inverse relationship between price and volume is not a reflection of market demand elasticity but an artifact of this two-phase declaration strategy, effectively visualizing the transition from value inflation to a realistic representation of a high-volume scrap recycling business.

**5.3 Synthesis: From Global Insight to Actionable Intelligence**

The convergence of the 390210 family validates that our framework can successfully translate macro-level red flags into a specific, HS Code-level watchlist. Furthermore, the anomaly detection algorithm provided a concrete, actionable insight by flagging a significant volume spike in the firm's top-value HS Code.

**Figure 28. Detailed Anomaly Detection for the Top-Value HS Code: 3921904200.**

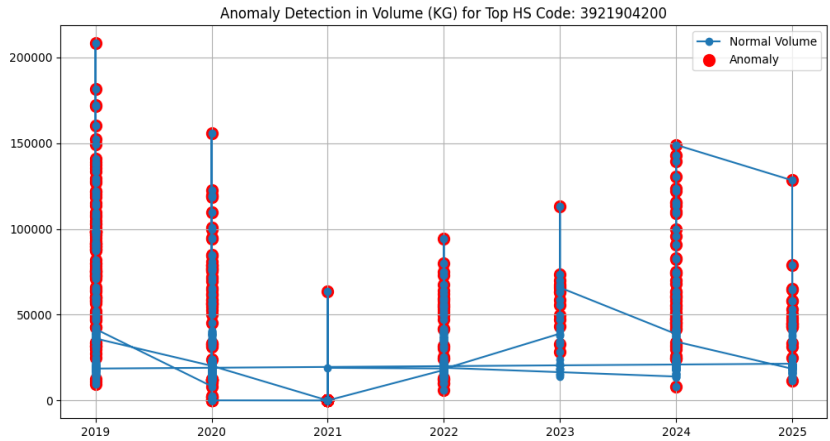

Source: Processed by Author (2025)

For a compliance officer in Malaysia, this is not a statistical curiosity it is an immediate directive. This finding justifies a transactional audit or deeper supply chain verification for that specific product to ensure compliance with Malaysia's strict regulations on waste importation. By pinpointing specific, high-risk transactions, the framework transforms from a descriptive tool into a predictive instrument for proactive due diligence, helping companies navigate complex regulatory environments and secure their supply chains against illicit trade.

**Figure 29. Anomaly Detection in Traded Volume for At-Risk Firm-Level HS Code.**

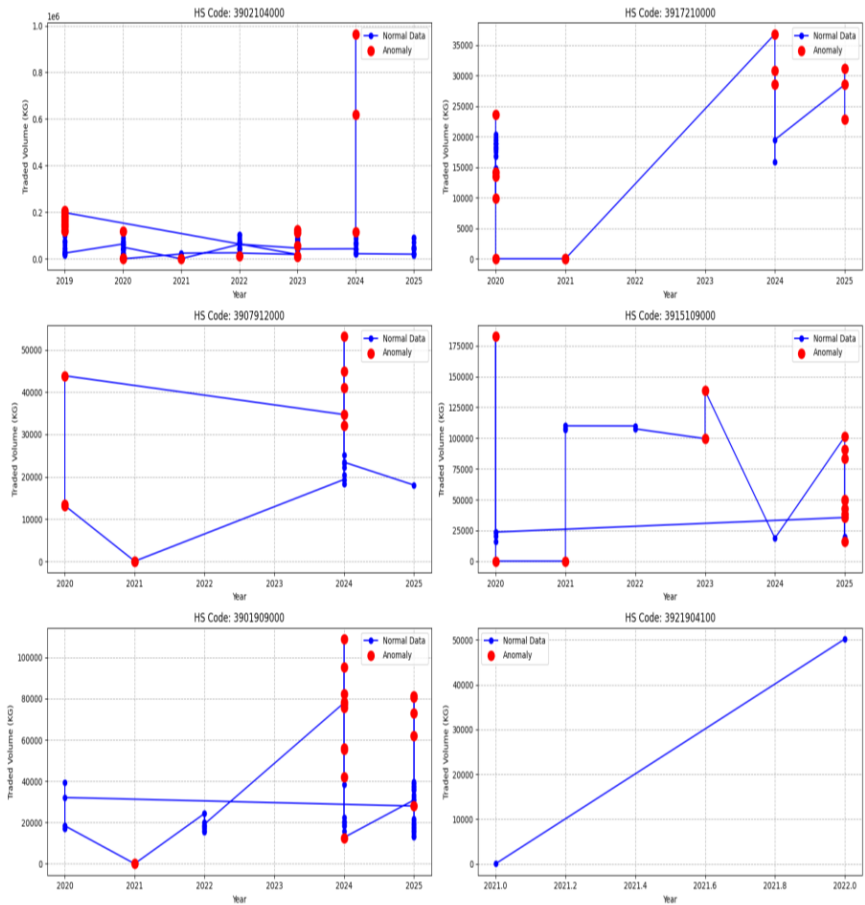

Source: Processed by Author (2025)

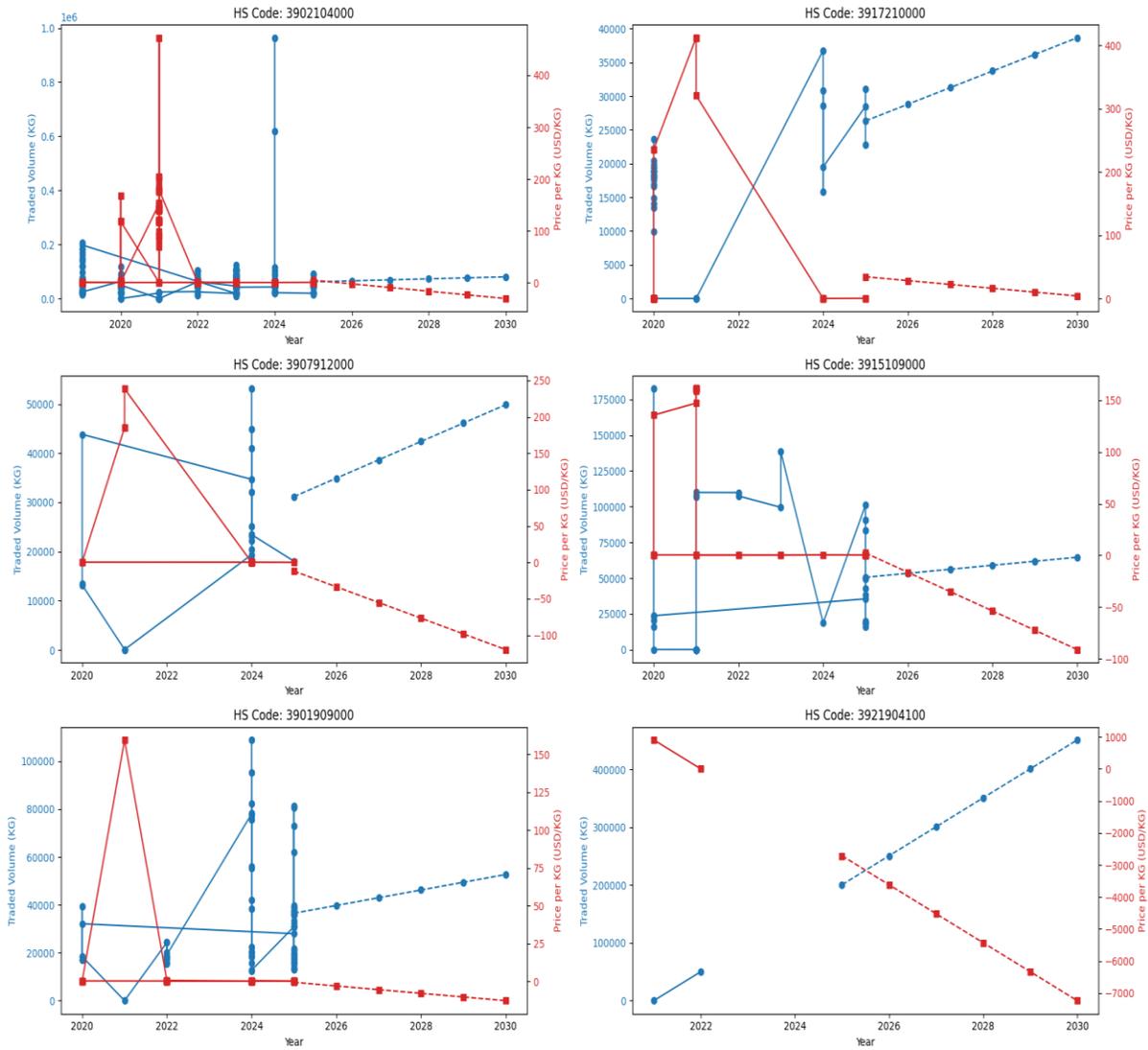

**Figure 30. Forecast of Trade Volume and Price per KG to 2030 for At-Risk Firm-Level HS 39.**

Source: Processed by Author (2025)

## 6. CONCLUSION

This study has demonstrated the feasibility and effectiveness of an interpretable, end-to-end machine learning framework for detecting the inverse price-volume signature of plastic-waste misclassification. By systematically integrating global UN Comtrade HS 39 data (2020-2024) with proprietary firm-level records (2019-2025), the framework achieved over ninety percent accuracy in both macro- and micro-scale evaluations. Central to its success is the identification of a consistent data marker rising reported volumes paired with collapsing unit prices that reliably flags illicit classification practices. The modelling pipeline, from K-Means market segmentation through Isolation Forest anomaly detection to Random Forest classification enriched by SHAP explainability, converts complex trade patterns into transparent, risk-based watchlists. These outputs empower customs authorities, such as the National Customs Department under its latest scrutiny of Certificate of Approval for HS 39.15 imports, and corporate compliance teams to prioritize inspections and mitigate high-risk flows before they materialize into regulatory breaches or environmental harm.

Our findings also highlight specific HS codes most notably 390210 and 392690 that warrant enhanced scrutiny and inclusion in international treaty provisions. To transition from proof-of-concept to operational enforcement, we propose a phased roadmap: first, pilot deployment of the model and officer training on a customs authorities test dashboard in Q4 2025; second, validation of flagged shipments against Port Klang seizure records in Q1 2026; third, calibration of anomaly and classification thresholds based on real-world outcomes in Q2 2026; and finally, scale-up of the system to additional HS 39 subcategories and partner countries by Q4 2026. Concurrently, negotiators of the international Plastics Treaty should explicitly list HS 390210 and HS 392690 under Annex categories requiring Prior Informed Consent, mandate cross-border data-sharing protocols for real-time anomaly detection, and incorporate regular model-audit requirements to maintain transparency, legal defensibility, and adaptability to evolving trade patterns. By following this integrated pathway, stakeholders can shift from reactive, seizure-driven enforcement to a proactive, data-driven regime that strengthens the integrity of global governance in plastic waste management.

Our findings illuminate a clear strategic path forward, beginning with the immediate operational deployment of this framework. We suggest integrating our risk-based watchlists into existing customs and compliance systems via a secure API, supported by robust protocols for threshold tuning and targeted officer training. Concurrently, theoretical research should aim to fortify the system against adversarial manipulation by developing advanced graph-based metrics and hybrid time-series models. The framework's ultimate potential, however, lies in its expansion beyond plastics to other high-risk commodities such as timber and electronic waste leveraging higher-frequency data and an automated alert dashboard to forge a harmonized, data-driven enforcement strategy across international jurisdictions.